\documentclass[prl,twocolumn,showpacs,preprintnumbers,amsmath,amssymb,superscriptaddress,floatfix]{revtex4-2}

\usepackage{amsfonts}
\usepackage{amsmath}
\usepackage{amssymb}
\usepackage{mathrsfs}
\usepackage{subfigure}
\usepackage{graphicx}
\usepackage{ulem}
\usepackage{epstopdf}
\usepackage{color}
\usepackage{bm}
\usepackage{braket}
\usepackage{makecell}
\usepackage{xcolor}
\usepackage{booktabs}
\usepackage{hyperref}
\usepackage{array}
\usepackage{placeins}

\hypersetup{colorlinks=true, linkcolor=blue, filecolor=blue, urlcolor=blue, citecolor=blue}
\def\equationautorefname~#1\null{%
	Eq.~#1\null
}
\def\figureautorefname~#1\null{%
	Fig.~#1\null
}

\begin{document}

\title{Finite Coherence in Gravitational Waves from Tidally Excited Axion Clouds}
\author{Yizhi Liang}
\affiliation{College of Physics, Sichuan University, Chengdu, 610065, China}
\author{Mian Zhu}
\affiliation{College of Physics, Sichuan University, Chengdu, 610065, China}
\author{Wen-Biao Han}
\affiliation{Shanghai Astronomical Observatory, CAS, 200030, Shanghai, China}
\affiliation{School of Fundamental Physics and Mathematical Sciences, Hangzhou Institute for Advanced Study, UCAS, Hangzhou 310024, China}
\author{Lianfu Wei}
\affiliation{Information Quantum Technology Laboratory, School of Information Science and Technology, Southwest Jiaotong University, Chengdu 610031, Sichuan, China}
\author{Peng Wang}
\affiliation{College of Physics, Sichuan University, Chengdu, 610065, China}
\author{Jun Tao}\email{taojun@scu.edu.cn}
\affiliation{College of Physics, Sichuan University, Chengdu, 610065, China}
\date{\today}

\begin{abstract}
	Axion clouds around rotating black holes form gravitational atoms whose tidal transitions can radiate gravitational waves in binaries. For strongly coupled Bohr crossings, transition radiation is governed by the outgoing two-level coherence, not by the transition probability alone. This coherence is suppressed both on the adiabatic branch and in the weak passage limit, but survives for intermediate sweep rates, producing a finite transition waveform and a localized orbital response. In more massive systems, fine and hyperfine transitions produce narrowband gravitational radiation and cumulative departures from vacuum binary waveforms. Coherent tidal crossings offer a gravitational-wave probe of axion-cloud dynamics.
\end{abstract}

\maketitle
{\it Introduction}---Ultralight axionlike particles are motivated by generic compactification spectra beyond the Standard Model \cite{Peccei:1977hh,Peccei:1977ur,Arvanitaki:2009fg,Marsh:2015xka}. Although their weak nongravitational couplings hinder laboratory searches, ultralight bosons can be probed through black hole superradiance: for Compton wavelengths comparable to the gravitational radius, a spinning black hole transfers rotational energy to the field, producing a macroscopic boson
cloud, often described as a gravitational atom \cite{Detweiler:1980uk,Arvanitaki:2010sy,Brito:2015oca}. In a binary, the companion's tidal field resonantly mixes cloud levels and exchanges energy and angular momentum with the orbit \cite{Arvanitaki:2010sy,Brito:2017zvb}. 
\par Recent studies of gravitational atoms have focused on fine and hyperfine transitions in slowly evolving systems, where this exchange accumulates as a long-lived phase drift \cite{Kyriazis:2025fis,Tomaselli:2024bdd,Tomaselli:2024dbw,DellaMonica:2025zby}.  Strongly coupled Bohr transitions behave differently. Previous work has shown that tidal resonances can transfer cloud occupation and, in some regimes, appreciably deplete the populated level \cite{Baumann:2018vus,Baumann:2019eav,Baumann:2021fkf,Boskovic:2024fga,Boskovic:2025ixx}. The radiative Bohr signal, however, is set by the interference term left after the crossing. In an adiabatic passage the cloud follows an instantaneous eigenstate, so the two-level coherence vanishes and the persistent transition quadrupole is quenched. This loss of the outgoing beat is the adiabatic quench considered below.

\par A Bohr crossing produces a persistent transition quadrupole only if the passage leaves both cloud levels populated with a definite relative phase. The Landau-Zener ratio \(z_{\rm LZ}\) compares the squared tidal coupling with the orbital sweep rate. Slow passages erase the outgoing beat, while very weak passages barely populate the second level. For intermediate \(z_{\rm LZ}\), the outgoing cloud carries finite two-level coherence and radiates a finite transition waveform. Combined with the superradiance and cloud overlap conditions, this criterion defines the region where a specified transition-strain threshold can be met. Lower frequency fine and hyperfine transitions occupy the long observation limit, where they produce narrowband radiation and accumulated waveform distortions \cite{Yagi:2011wg,Kapil:2024zdn,Vicente:2025gsg,Cheng:2025wac}.

{\it Framework}---Kerr black holes support hydrogenic axion clouds when the superradiance condition $0<\omega<m\Omega_H$ is satisfied \cite{Brito:2015oca,Marsh:2015xka}. For an axion of mass $\mu$, the fine-structure constant is $\alpha=\mu M_1$ \cite{Brito:2017zvb}, where $M_1$ is the mass of the central black hole. A companion of mass $M_*=M_2$ on an eccentric, equatorial orbit perturbs the cloud through
a multipolar tidal potential whose matrix element is written as $V_{ge}(t)=\sum_n\eta_n(e_{\rm orb},a_{\rm orb})e^{-in\Phi(t)}$. The states $|g\rangle$ and $|e\rangle$ denote the levels populated before and after the resonant crossing; \(c_g\) and \(c_e\) are their amplitudes, and \(E_g\) and \(E_e\) are their level angular frequencies. The two-level equations are written in the natural unit convention in which \(E_i\), \(\Delta E\), \(\eta_n\), and \(\Gamma\) are angular frequencies; the corresponding physical energy is \(\hbar E_i\). The coupling $\eta_n$ contains the radial cloud overlap, the Clebsch-Gordan angular factor, and the eccentric harmonic coefficient of the $n$th orbital harmonic; in the far-field limit this reduces to the usual Hansen coefficient, while the compact Bohr examples retain the variation of the cloud overlap along the eccentric orbit. \(a_{\rm orb}\), \(e_{\rm orb}\), and \(\Phi(t)\) are the semimajor axis, eccentricity, and mean orbital phase. In the equatorial geometry, the angular integral enforces $\Delta m=m_\ast$, $|l_1-l_2|\le l_\ast\le l_1+l_2$, and $(-1)^{l_1+l_2}=(-1)^{\Delta m}$; eccentricity turns the tidal field into a harmonic comb, and the $n$th harmonic resonates when $\Delta E\simeq n\Omega_{\rm orb}$, with $\Delta E=E_e-E_g$. Near an isolated resonance, extracting the slowly varying envelope by $\tilde c_e=c_e e^{in\Phi}$ gives the two-level cloud equation \cite{Kyriazis:2025fis,Baumann:2018vus,Baumann:2019eav,Lyu:2026mbv,Kim:2025wwj}
\begin{equation}
	i\frac{d}{dt}\begin{pmatrix} c_g(t) \\ c_e(t) \end{pmatrix} = \begin{pmatrix} E_g & \sum_n \eta_n e^{-in\Phi(t)} \\ \sum_n \eta_n^* e^{in\Phi(t)} & E_e - i\Gamma \end{pmatrix} \begin{pmatrix} c_g(t) \\ c_e(t) \end{pmatrix}
\end{equation}
where $\Gamma$ is the superradiance growth or damping rate obtained from the complex cloud eigenfrequency. The same coherent transition rate $\mathcal{R}(t)=-2\Im(\eta\,c_g^*\tilde c_e)$ modifies the orbital evolution, with \(\eta\) denoting the resonant harmonic \(\eta_n\). Defining $\Delta\omega=\omega_e-\omega_g$, $\Delta m=m_e-m_g$, $M_{\rm tot}=M_1+M_2$, $\mu_{\rm red}=M_1M_2/M_{\rm tot}$, and cloud mass \(M_c\), the semimajor axis \(a\) and eccentricity \(e\) obey
\begin{equation}
	\begin{split}
		\frac{da}{dt}&= \left( \frac{da}{dt} \right)_{P}
		-\frac{2a^2 M_c \Delta\omega}{\alpha c M_2}\mathcal{R}(t),\\
		\frac{de}{dt} &= \left( \frac{de}{dt} \right)_{P}
		- \frac{1-e^2}{e}\,\mathcal{R}(t)\\
		&\quad \times
		\left[ \frac{a M_c \Delta\omega}{\alpha c M_2}
		- \frac{G M_1 M_c \Delta m}
		{\alpha c\,\mu_{\rm red} \sqrt{G M_{\rm tot}a(1-e^2)}} \right],
	\end{split}
\end{equation}
where the subscript $P$ denotes the leading vacuum quadrupole radiation reaction terms \cite{Peters:1964zz}. Upward transitions absorb orbital energy into the cloud; downward transitions release stored cloud energy and can slow the inspiral, producing a floating response. The two signs imprint different orbital responses while being governed by the same coherent transition rate. Near an isolated crossing the local Landau-Zener description applies \cite{Landau:1932vnv,Landau:1932wdt}
with detuning $\delta_n(t)=n\Omega_{\rm orb}(t)-\Delta E$ and $\dot\delta_n\simeq n\dot\Omega_{\rm orb}$ evaluated at resonance. The dimensionless ratio \(z_{\rm LZ}\) compares the squared tidal coupling with the resonance sweep rate and determines whether the crossing leaves a coherent transition signal,
\begin{equation}
	z_{\rm LZ} = \frac{|\eta_n|^2}{|\dot\delta_n|}
	\simeq \frac{|\eta_n|^2}{n\dot{\Omega}_{\rm orb}},
	\label{eq:lz_parameter}
\end{equation}
Starting from a populated cloud level, the outgoing Landau-Zener scattering matrix gives
\begin{equation}
	P_{\rm tr}=1-e^{-2\pi z_{\rm LZ}},\qquad
	\mathcal{C}_{\rm out}=|c_g^\ast c_e|_{\rm out}
	=\sqrt{P_{\rm tr}(1-P_{\rm tr})}.
	\label{eq:cout}
\end{equation}
The transition waveform is controlled by the outgoing coherence \(\mathcal C_{\rm out}\). In the adiabatic limit \(z_{\rm LZ}\gg1\), the cloud population follows the instantaneous eigenstate; in the cloud depletion cases studied previously this transfers or removes the original cloud population \cite{Baumann:2018vus,Takahashi:2021yhy,Tomaselli:2024dbw}, while the asymptotic two-level beat satisfies \(\mathcal C_{\rm out}\to0\). The weak passage limit \(z_{\rm LZ}\ll1\) also gives \(\mathcal C_{\rm out}\to0\). A finite transition waveform appears in the intermediate regime where both outgoing levels remain populated. \(\mathcal C_{\rm out}\) is nonmonotonic in \(z_{\rm LZ}\), reaches its maximum value \(1/2\) at \(z_{\rm LZ}=\ln 2/(2\pi)\), and is exponentially suppressed on the adiabatic branch. The Bohr transition waveform is therefore governed by the sweep parameter \(z_{\rm LZ}\).
\par The waveform is conditional on a populated cloud level satisfying \(\omega_{nlm}<m\Omega_H\) at the resonance. The cloud mass \(M_c\) fixes the strain normalization, while the prior growth, depletion, and possible self-interaction history determine the available occupation. In the compact Bohr example the orbit crosses the radial support of the cloud, giving finite cloud-orbit overlap during the resonance.

\begin{figure}[tbp]
	\centering
	\includegraphics[width=\linewidth]{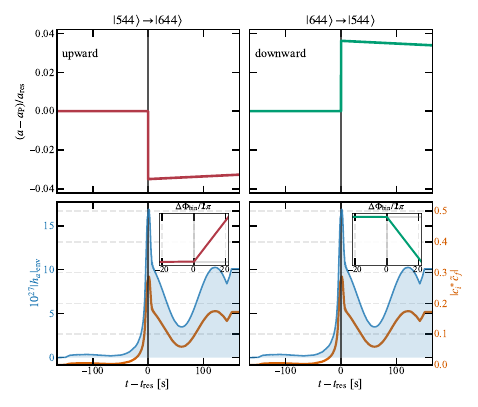}
	\caption{High-frequency Bohr events for the $\Delta m=0$ transitions $|544\rangle\rightarrow|644\rangle$ (left) and $|644\rangle\rightarrow|544\rangle$ (right), plotted relative to the resonance. The parameters are $M_1=10^{-2}\,M_\odot$, $q=M_2/M_1=10^{-3}$, $\alpha=0.30$, $a_\ast=0.70$, $e_0=0.65$, $d_L=1\,{\rm kpc}$, and $M_c/M_1=10^{-4}$. Top: shift in semimajor axis relative to the vacuum binary evolution. Bottom: blue shading denotes the smoothed transition-strain envelope, and the orange curve denotes the cloud coherence \( |c_i^\ast\tilde c_f| \) on the right axis. Insets show the binary phase residual $\Delta\Phi_{\rm bin}/2\pi$ relative to the vacuum PN template, making the opposite upward and downward backreaction signs explicit. The vertical line marks the resonance.}
	\label{fig:bohr_pair}
\end{figure}

{\it Bohr crossings with finite outgoing coherence}---For Bohr transitions such as $|644\rangle\rightarrow|544\rangle$, the tidal matrix element is large. In slowly evolving binaries this can place the crossing on the adiabatic branch and quench the interference term $c_g^\ast\tilde c_e$ that radiates gravitational waves from axion transitions. In compact binaries the same crossing family can instead lie in the intermediate \(z_{\rm LZ}\) regime, where the outgoing state is neither fully transferred nor nearly unchanged. For the $\Delta m=0$ Bohr pair shown in Fig.~\ref{fig:bohr_pair}, the axisymmetric transition quadrupole gives
\begin{equation}
	\begin{split}
		h_{a,+}^{(0)}(t) &= -\mathcal{A}_0 \sin^2\iota\,\mathcal{C}(t),\\
		h_{a,\times}^{(0)}(t) &= 0,
	\end{split}
\end{equation}
where
\begin{equation}
	\mathcal{C}(t)=\Re[c_g^*(t)\tilde{c}_e(t)]\cos(\Delta E\,t)-\Im[c_g^*(t)\tilde{c}_e(t)]\sin(\Delta E\,t),
\end{equation}
	and $\mathcal{A}_0 = \frac{4G}{c^4 d_L} M_c (r_g/\alpha^2)^2 \omega_\text{trans}^2 \mathcal{F}_{\Delta m=0}$. Here \(d_L\) is the luminosity distance, \(r_g=GM_1/c^2\) is the gravitational radius, \(\omega_{\rm trans}=|\Delta E|\), and \(\mathcal{F}_{\Delta m=0}\) is the dimensionless transition-quadrupole factor. The inclination \(\iota\) is measured from the cloud spin axis, and the \(\sin^2\iota\) term is the axisymmetric quadrupole pattern. The Bohr examples report the projected strain with \(\sin^2\iota=1\). For this conjugate \(\Delta m=0\) pair the transition-strain envelope is nearly independent of transition direction, since the two transitions share \(|\Delta E|\) and the same transition quadrupole. The sign of the energy exchange instead appears in the orbital response, which gives opposite phase residuals. For the \(\Delta m=2\) fine and hyperfine transitions used below, the same coherence factor \(\mathcal{C}(t)\) is projected with the usual quadrupolar factors \(h_{a,+}\propto(1+\cos^2\iota)\mathcal{C}/2\) and \(h_{a,\times}\propto\cos\iota\,\mathcal{C}\).
\par The time-domain waveform separates the cloud radiation from the orbital backreaction. At the resonant crossing, the cloud-orbit energy exchange appears as an abrupt offset in the semimajor axis relative to vacuum evolution; after the passage, the cloud envelope settles to a nonzero post-crossing plateau only if both levels remain populated. The conjugate transitions have nearly the same transition-strain envelope because they share \(|\Delta E|\) and the same transition quadrupole, while the jump in semimajor axis and binary phase residual reverse sign with the direction of cloud energy exchange. The envelope measures the surviving two-level coherence, whereas the orbital residual records whether the cloud absorbs or releases orbital energy.
\par
\begin{figure}[tbp]
	\centering
	\includegraphics[width=\linewidth]{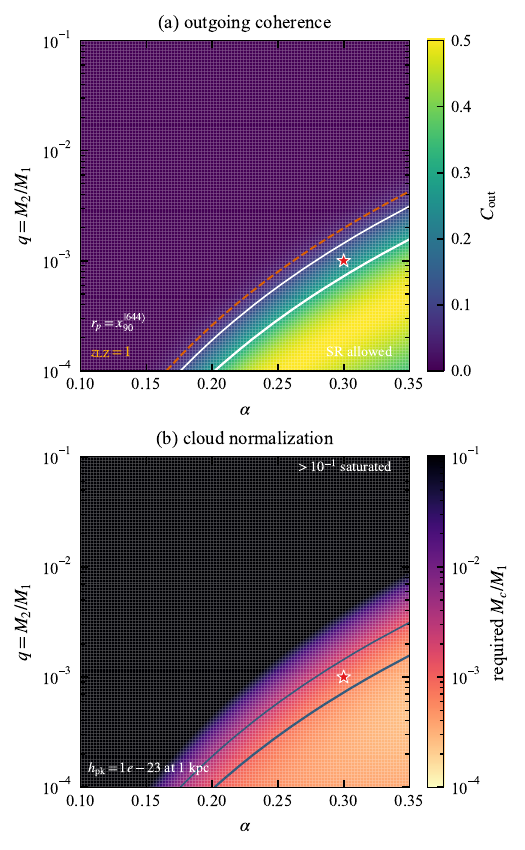}
	\caption{Domain estimate for the rapidly swept Bohr crossing. (a) Outgoing coherence \(C_{\rm out}\) in the \((\alpha,q)\) plane for the \(n=1\), \(|644\rangle\leftrightarrow|544\rangle\) crossing at fixed \(e_{\rm res}=0.64\) and \(a_\ast=0.70\). White contours mark \(C_{\rm out}=0.1\) and \(0.3\), the dashed orange contour marks \(z_{\rm LZ}=1\), and the dotted contour marks \(r_p=x_{90}^{|644\rangle}\). The star denotes the event shown in Fig.~\ref{fig:bohr_pair}. (b) Cloud mass fraction required to reach \(h_{\rm pk}=10^{-23}\) at \(d_L=1\,{\rm kpc}\). Values above \(M_c/M_1=0.1\) are saturated in the color scale. The two panels identify the domain of coherent outgoing transition signals and the associated strain scale.}
	\label{fig:domain_visibility}
\end{figure}

 Figure~\ref{fig:domain_visibility} maps the domain of Bohr crossings with finite outgoing coherence. Panel (a) displays the outgoing coherence at fixed eccentricity, together with the \(z_{\rm LZ}=1\) contour, the superradiance condition for the participating level, and a compactness contour comparing the resonant pericenter with the radial support of the \(|644\rangle\) cloud. Panel (b) reports the cloud mass fraction required to obtain \(h_{\rm pk}=10^{-23}\) at \(d_L=1\,{\rm kpc}\), using the linear dependence of strain on \(M_c\). The star marks the case shown in Fig.~\ref{fig:bohr_pair}. Together, the two panels specify, in the same \((\alpha,q)\) plane, the conditions for coherent passage, cloud existence, compact orbital overlap, and the chosen strain scale.
\par
\begin{figure}[tbp]
	\includegraphics[width=\linewidth]{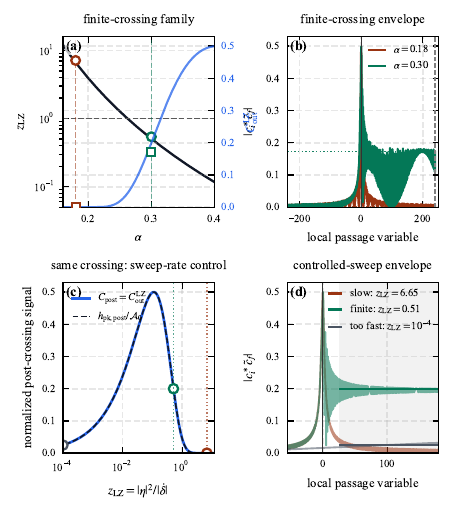}
	\caption{Criterion for finite outgoing coherence in Bohr transition waveforms. (a) For the \(|644\rangle\rightarrow|544\rangle\) family, changing \(\alpha\) maps the crossing onto \(z_{\rm LZ}\) and the analytic outgoing coherence. (b) Direct local passages show that the post-crossing envelope is suppressed on the adiabatic branch but remains finite near the event in Fig.~\ref{fig:bohr_pair}. (c) Varying the sweep rate at fixed strain scale gives \(h_{\rm pk,post}/\mathcal A_0=C_{\rm out}\). (d) The corresponding local envelopes show suppression in both the adiabatic and weak passage limits, with finite coherence in between.}
	\label{fig:bohr_lz_evidence}
\end{figure}

 Figure~\ref{fig:bohr_lz_evidence} tests the coherence criterion by comparing the analytic \(C_{\rm out}(z_{\rm LZ})\) relation with direct local passages of the eccentric \(q=10^{-3}\), \(e_{\rm res}\simeq0.64\), \(n=1\), \(|644\rangle\rightarrow|544\rangle\) family. At \(\alpha=0.18\), the passage is on the adiabatic branch, \(z_{\rm LZ}=7.12\), and the residual coherence is strongly suppressed, \(C_{\rm post}=2.0\times10^{-4}\). At the \(\alpha=0.30\) point used in Fig.~\ref{fig:bohr_pair}, \(z_{\rm LZ}=0.541\), both levels remain populated, and \(C_{\rm post}=0.172\). The right panels vary the sweep rate at fixed strain scale, showing that waveform visibility follows the nonmonotonic outgoing coherence rather than the monotonic transition probability. Finite Bohr radiation is concentrated at intermediate \(z_{\rm LZ}\).

\begin{figure}
	\centering
	\includegraphics[width=\linewidth]{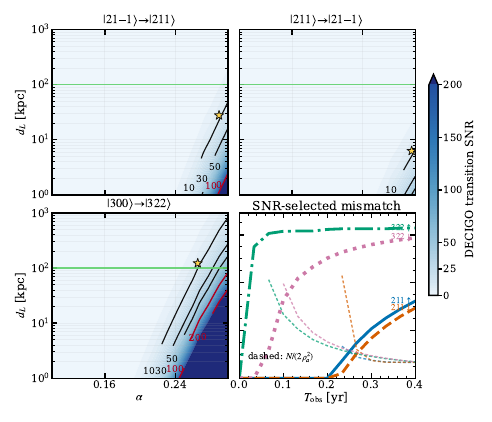}
	\caption{Low-frequency resolved source signatures for $M_1=1500M_\odot$, $M_2=0.5M_\odot$, $e_0=0.30$, $M_c/M_1=0.05$, reference distance $d_L=100\,{\rm kpc}$, and a $0.4\,{\rm yr}$ observation. The first three panels show projected transition signal SNR for DECIGO design sensitivity, including the four-unit DECIGO response normalization, for transitions whose direct SNR exceeds threshold; stars mark the weakest retained direct detections, with $\rho_a\gtrsim8$, used in the mismatch comparison. The fourth panel shows the DECIGO-noise-weighted mismatch between the waveform including cloud backreaction and a vacuum PN template for those configurations; dashed curves denote the cloud-component perturbative scale \(N/(2\rho_a^2)\) with \(N=13\). For the upward cases the resonant \(n=4\) passage lies inside the observation window; the downward curves show accumulated waveform deformation because their direct transition SNR is below threshold. Early flat segments in the hyperfine curves precede the accumulated resonance crossing.}
	\label{fig:lowfre_diag}
\end{figure}

{\it Low-frequency resolved signatures}--- In more massive systems the same two-level dynamics is probed in a different observational limit: the cloud radiation is narrowband over the observing window, and the backreaction is tested through accumulated waveform differences. Fine and hyperfine resonances persist long enough to form narrowband transition features, and their backreaction also shifts the binary waveform away from a vacuum inspiral. The resolved source comparison uses two diagnostics: the transition-strain SNR, evaluated with the four-unit DECIGO response normalization \cite{Kyriazis:2025fis,Yagi:2011wg}, and the noise-weighted mismatch between the waveform including cloud backreaction \(h_{\rm total}=h_{\rm bin}+h_a\) and a fixed vacuum PN template \cite{Dai:2023cft,Zhao:2024bpp},
\begin{equation}
	\mathcal{M}_{mis}[h_t, h_p] = 1 - \max_{t_0, \phi_0} \frac{\langle h_t | h_p e^{i(2\pi f t_0 + \phi_0)}\rangle}{\sqrt{\langle h_t|h_t\rangle \langle h_p|h_p\rangle}}.
\end{equation}
The mismatch is assessed against the cloud-component perturbative scale \(N/(2\rho_a^2)\) \cite{Robson:2018ifk,Babak:2006uv,Lindblom:2008cm}, where \(N=13\) and \(\rho_a\) is the transition SNR of the cloud signal in the same detector band. This scale is set by the cloud signal itself, rather than by the larger SNR of the vacuum binary foreground. Figure~\ref{fig:lowfre_diag} compares the SNR and mismatch for the same resonant configurations. For the parameters in the caption, the direct transition signal has \(\rho_a\simeq82\) for \(|300\rangle\to|322\rangle\) and \(\rho_a\simeq4.9\) for \(|21{-}1\rangle\to|211\rangle\), while the downward transitions lie below the SNR threshold. For the weakest retained direct detections, \(\rho_a\gtrsim8\), the upward transitions cross the resonant \(n=4\) harmonic during the observation. The downward transitions appear through accumulated waveform deformation because their direct transition SNR is below threshold. The resulting mismatches, \(\mathcal{M}_{mis}=0.44\)--\(0.94\), exceed the corresponding cloud-component scales \(N/(2\rho_a^2)\simeq0.10\). In this regime the axion cloud remains a perturbation to the binary, while producing both cloud radiation and a noise-weighted departure from a vacuum waveform.

{\it Discussion and Conclusion}---Bohr transition radiation is controlled by the outgoing two-level coherence \( |c_g^\ast c_e|_{\rm out} \), not by the transition probability alone. The signal is quenched on the adiabatic branch, \(z_{\rm LZ}\gg1\), where the post-crossing beat vanishes, and it is also weak for \(z_{\rm LZ}\ll1\), where the second level is scarcely populated. An eccentric compact crossing with intermediate \(z_{\rm LZ}\) can instead leave a coherent two-level cloud and radiate a finite transition waveform. Bohr radiation is distinct from cloud depletion and phase drift observables: the transition quadrupole is sourced by the surviving interference between outgoing cloud levels. The accompanying orbital impulse changes sign depending on whether the cloud absorbs or releases orbital energy. The domain map specifies where superradiance, compact orbital overlap, and the cloud occupation required for a given strain can be satisfied simultaneously. In the long observation limit, the lower frequency fine and hyperfine examples appear through narrowband features and accumulated waveform distortions.
\par This signal requires the relevant superradiant level to be populated at resonance; cloud formation, nonlinear depletion, and event rates depend on the astrophysical formation history. If small black holes with large spins and axion clouds occur in dense environments, rapidly swept transitions contribute to stochastic background estimates through the mapping from an individual event spectrum to \(\Omega_{\rm GW}\) \cite{Phinney:2001di,Regimbau:2011rp,Rosado:2011kv}. Potential source populations include compact objects retained in dark-matter overdensities, dense stellar clusters, nuclear environments, and AGN disks \cite{Eda:2014kra,Rodriguez:2016kxx,Antonini:2016gqe,Stone:2016wzz,Yang:2019bqg}. More speculative scenarios involving primordial black holes with large spins, including formation during an early matter dominated era, are another possible source class \cite{Harada:2017fjm,deJong:2023gsx}.

{\it Acknowledgments}---The authors are grateful to Guohe Li, Ran Chen and Rui Yang for useful discussions and insightful suggestions. This work is supported by the National Science and Technology Major Project (No. 2024ZD1100601), National Natural Science Foundation
of China (NSFC) with Grants Nos. 12175212, 12275183, 12473075, 12503005, Sichuan Science and Technology Program Grant No. 2026NSFSC0804, and the Fundamental Research Funds for the Central Universities Grant No. YJ202551.

\bibliography{r}

\end{document}